\documentclass[prl,twocolumn,superscriptaddress]{revtex4-2}
\usepackage{amssymb}
\usepackage{amsfonts}
\usepackage{mathrsfs}
\usepackage{graphics,graphicx,epsfig,amsmath,amsthm}
\usepackage{floatrow}
\usepackage{caption}
\usepackage[labelformat=simple]{subfig}

\usepackage{upgreek}
\usepackage{bm}
\usepackage{bbm}
\usepackage{multirow}
\usepackage{array}
\usepackage{color}
\usepackage{cases}
\usepackage{booktabs }
\usepackage[usenames,dvipsnames]{xcolor}

\usepackage{xcolor}

\usepackage{float}
\usepackage[colorlinks=true,linkcolor=black,citecolor=blue,urlcolor=blue]{hyperref}
\usepackage[doipre={doi:~}]{uri}
\makeatletter
\let\saved@includegraphics\includegraphics
\AtBeginDocument{\let\includegraphics\saved@includegraphics}
\renewenvironment*{figure}{\@float{figure}}{@float}
\makeatother
\begin{document}

\title{Experimental study on the principle of minimal work fluctuations}
\author{Wei Cheng}
\affiliation{CAS Key Laboratory of Microscale Magnetic Resonance and School of Physical Sciences, University of Science and Technology of China, Hefei 230026, China}
\affiliation{CAS Center for Excellence in Quantum Information and Quantum Physics, University of Science and Technology of China, Hefei 230026, China}
\author{Wenquan Liu}
\affiliation{School of Science, Beijing University of Posts and Telecommunications, Beijing, 100876, China}
\author{Yang Wu}
\affiliation{CAS Key Laboratory of Microscale Magnetic Resonance and School of Physical Sciences, University of Science and Technology of China, Hefei 230026, China}
\affiliation{CAS Center for Excellence in Quantum Information and Quantum Physics, University of Science and Technology of China, Hefei 230026, China}
\author{Zhibo Niu}
\affiliation{CAS Key Laboratory of Microscale Magnetic Resonance and School of Physical Sciences, University of Science and Technology of China, Hefei 230026, China}
\affiliation{CAS Center for Excellence in Quantum Information and Quantum Physics, University of Science and Technology of China, Hefei 230026, China}
\author{Chang-Kui Duan}
\affiliation{CAS Key Laboratory of Microscale Magnetic Resonance and School of Physical Sciences, University of Science and Technology of China, Hefei 230026, China}
\affiliation{CAS Center for Excellence in Quantum Information and Quantum Physics, University of Science and Technology of China, Hefei 230026, China}
\author{\\Jiangbin Gong}
\email{phygj@nus.edu.sg}
\affiliation{Department of Physics, National University of Singapore, Singapore 117551, Singapore}
\affiliation{Center for Quantum Technologies, National University of Singapore,
Singapore 117543, Singapore}
\affiliation{Joint School of National University of Singapore and Tianjin
University, International Campus of Tianjin University, Binhai New City,
Fuzhou 350207, China}
\affiliation{MajuLab, CNRS-UCA-SU-NUS-NTU International Joint Research Unit,
Singapore}
\author{Xing Rong}
\email{xrong@ustc.edu.cn}
\affiliation{CAS Key Laboratory of Microscale Magnetic Resonance and School of Physical Sciences, University of Science and Technology of China, Hefei 230026, China}
\affiliation{CAS Center for Excellence in Quantum Information and Quantum Physics, University of Science and Technology of China, Hefei 230026, China}
\affiliation{Hefei National Laboratory, University of Science and Technology of China, Hefei 230088, China}
\author{Jiangfeng Du}
\email{djf@ustc.edu.cn}
\affiliation{CAS Key Laboratory of Microscale Magnetic Resonance and School of Physical Sciences, University of Science and Technology of China, Hefei 230026, China}
\affiliation{CAS Center for Excellence in Quantum Information and Quantum Physics, University of Science and Technology of China, Hefei 230026, China}
\affiliation{Hefei National Laboratory, University of Science and Technology of China, Hefei 230088, China}
\affiliation{School of Physics, Zhejiang University, Hangzhou 310027, China}

\begin{abstract}
The central quantity in the celebrated quantum Jarzynski equality is $e^{-\beta W}$, where $W$ is work and $\beta$ is the inverse temperature. The impact of quantum randomness on the fluctuations of $e^{-\beta W}$ and hence on the predictive power of the Jarzynski estimator is an important problem.  Working on a single nitrogen-vacancy center in diamond and riding on an implementation of two-point measurement of non-equilibrium work with single-shot readout,  we have conducted a direct experimental investigation of the relationship between the fluctuations of $e^{-\beta W}$ and adiabaticity of non-equilibrium work protocols.  It is observed that adiabatic processes minimize the variance of $e^{-\beta W}$, thus verifying an early theoretical concept, the so-called principle of minimal work fluctuations.  Furthermore, it is experimentally demonstrated that shortcuts-to-adiabaticity control can be exploited to minimize the variance of  $e^{-\beta W}$ in fast work protocols.  Our work should stimulate further experimental studies of quantum effects on the bias and error in the estimates of free energy differences based on the Jarzynski equality.  
\end{abstract}
\maketitle

\begin{figure*}
\centering
\includegraphics[width=0.95\columnwidth]{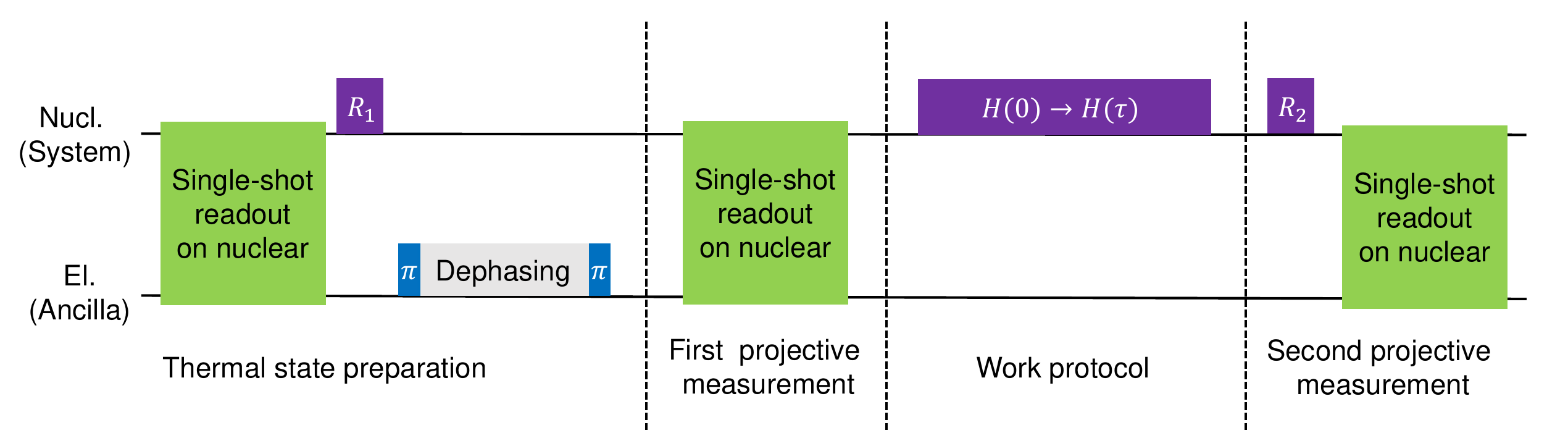}
\caption{\label{fig1} Experimental pulse sequence to study the PMWF in the NV center platform. The pulse sequence is used to obtain work statistics via the two-point measurement scheme, which includes the thermal state preparation, the first projective measurement, the work protocol and the second projective measurement.
}
\end{figure*}

\section{I. Introduction}

Thermal and quantum fluctuations are dominant features in non-equilibrium thermodynamics~\cite{Esposito2009,Campisi2011,Seifert2012,Hanggi2015,Funo2018} and can be quantified by fluctuation theorems~\cite{Jarzynski1997,Crooks1999,Mukamel2003,Talkner2007,Campisi2009,Sagawa2010,An2015}. Fluctuation theorems are not only insightful for our understanding of the second law of thermodynamics but also instrumental for us to extract equilibrium thermodynamic quantities from non-equilibrium processes~\cite{Liphardt2002,Harris2007,Gupta2011,Raman2014}.  
The celebrated Jarzynski equality~\cite{Jarzynski1997} does apply to quantum systems and assumes essentially the same form as in the classical case, namely, $\langle e^{-\beta W}\rangle = e^{-\beta\Delta F}$, where $\langle \cdot \rangle$ represents both thermal and quantum ensemble averaging. This equality connects the mean of $e^{-\beta W}$ with the exponential function of a free-energy difference $\Delta F$ at the inverse temperature $\beta$.  Interestingly, though suppression of work fluctuations is of great interest in the context of nanoscale heat engines~\cite{Deng2013,Deng2018,Peterson2019}, there is no experimental effort focusing on 
the fluctuations in the quantity $e^{-\beta W}$ measured in quantum processes. Such fluctuations can be enhanced by quantum randomness and are hence of crucial importance when applying the Jarzynski relation to quantum settings. 

For a finite number of measured work values, the Jarzynski free-energy estimator of $\Delta F$ is given by $-\beta^{-1}\ln[\sum_{k=1}^{N} e^{-\beta W_k}/N]$, where $W_k$ denotes work in the $k$th experimental run.  This estimator is a biased one, so when regarding practical applications the fluctuations of $e^{-\beta W}$ are crucial. It determines whether the variance of $e^{-\beta W}$ diverges, in which case the convergence of our estimate of $\Delta F$ will be extremely slow.  If the variance of $e^{-\beta W}$ does converge, it still determines the number of experimental runs ($N$) needed to reach a given estimate precision of $\Delta F$~\cite{Gore2003}.
Several theoretical studies~\cite{Xiao2014,Xiao2015,Deng2017,Jaramillo2017} touched upon the suppression of the fluctuations in $e^{-\beta W}$.  In particular, the so-called principle of minimal work fluctuations (PMWF)~\cite{Xiao2015} states that for a quantum system initially at thermal equilibrium but detached from its surrounding afterwards, the minimal variance of $e^{-\beta W}$ can be achieved in an adiabatic process if the instantaneous energy levels do not cross during the work protocol. This indicates if the variance diverges for the adiabatic process, then it diverges for all work protocols.
Adiabatic processes call for a long evolution time, however, the coherence time of realistic quantum systems is limited.
Regarding this, the PMWF also proved that one can equally use shortcuts-to-adiabatic (STA) processes~\cite{Berry2009,Chen2010} to achieve the minimal variance of $e^{-\beta W}$.
PMWF can thus guide us in the designing of work protocols in order to suppress the variance of $e^{-\beta W}$ and hence reduce experimental efforts when applying the Jarzynski equality to estimate $\Delta F$.   The suppression of the variance of  $e^{-\beta W}$ also naturally leads to the suppression of the fluctuations in the work itself~\cite{Xiao2014}. 

Here we report an experimental investigation of the PMWF utilizing a single nuclear spin of a nitrogen-vacancy ($\rm{NV}$) center in diamond.
Our study focuses on the variance of $e^{-\beta W}$ during a work protocol applied to a thermal equilibrium state.
The work statistic is directly obtained by the two-point measurement of work~\cite{Esposito2009,Campisi2011}. The realization of the two-point measurement of work makes it possible to examine the fluctuations in $e^{-\beta W}$ and hence verify the PMWF. 
Experimental results show that the variance of $e^{-\beta W}$ decreases gradually as the work protocol approaches the adiabatic limit.
Furthermore, we demonstrate that the variance of $e^{-\beta W}$ in the STA process, which is achieved by implementing a counter-diabatic (CD) driving Hamiltonian, is the same as that in adiabatic work protocols.   
Besides, the role of temperature on the variance of $e^{-\beta W}$ is also examined. It is observed that a higher-temperature setting suppresses the variance of $e^{-\beta W}$.  

\section{II. Theory}

We investigate the variance of $e^{-\beta W}$ when a system initially at thermal equilibrium undergoes a non-equilibrium work protocol while being isolated from the heat bath. During the protocol, an external control was applied to tune the system Hamiltonian from $H(0)$ to $H(\tau)$. The work done on the system can be obtained via the two-point measurement scheme, which requires performing two projective measurements at the beginning and the end of the protocol. 
This allows us to determine the trajectory, such as $|E_m(0)\rangle\to|\tilde{E}_n(\tau)\rangle$, in each realization of the work protocol, with $|E_m(0)\rangle$ and $|\tilde{E}_n(\tau)\rangle$ denoting the eigenstates of $H(0)$ and $H(\tau)$, respectively.
The work associated with the trajectory $|E_m(0)\rangle\to|\tilde{E}_n(\tau)\rangle$ is $W_{mn}=\tilde{E}_n-E_m$, where $E_m$ and $\tilde{E}_n$ are eigenvalues corresponding to $|E_m(0)\rangle$ and $|\tilde{E}_n(\tau)\rangle$, respectively.
The work distribution is given by 
\begin{equation}
P(W)=\sum_{m,n}p^0_{m}p^{\tau}_{m\rightarrow n}\delta[W-(\tilde{E}_n-E_m)] ,
\label{work_distribution}
\end{equation}
where $p^0_m$ is the initial thermal population of $|E_m(0)\rangle$ and $p^{\tau}_{m\rightarrow n}$ is the transition probability to $|\tilde{E}_n(\tau)\rangle$ conditioned that the trajectory started from $|E_m(0)\rangle$.
The variance of $e^{-\beta W}$ is directly obtained from
\begin{equation}
\sigma^2(e^{-\beta W})=\sum_WP(W)e^{-2\beta W}-\left[\sum_WP(W)e^{-\beta W}\right]^2 .
\label{work_fluctuation}
\end{equation}
The PMWF predicts that $\sigma^2(e^{-\beta W})$ achieves its minimal value when $p^{\tau}_{m\rightarrow n}=\delta_{mn}$, which can be realized in an adiabatic process or an STA process.

\section{III. Experiments}

In our experiment, a single nuclear spin of an $\rm{NV}$ center system with natural abundance of ${}^{13}$C (1.1$\%$) is used. 
The $\rm{NV}$ center is a type of defect in diamond which consists of a substitutional nitrogen atom adjacent to a carbon vacancy. 
The vacancy contains a spin-1 electron spin which couples to the spin-1 nuclear spin of the adjacent ${}^{14}$N atom.
Hamiltonian of the NV center system can be written as $H_{\rm{NV}} = 2\pi(DS_{z}^{2}+\omega_{e}S_{z}+QI_{z}^{2}+\omega_{n}I_{z}+A_{zz}I_{z}S_{z})$, where $S_{z}$ and $I_{z}$ are the spin operators of the electron spin and the nuclear spin, respectively.
Here, $D = 2.87\,\rm{GHz}$ is the ground-state zero-field splitting of the electron spin, $Q = -4.95\,\rm{MHz}$ is the quadrupolar interaction of nuclear spin, and $A_{zz} = -2.16\,\rm{MHz}$ is the longitudinal component of the hyperfine interaction between nuclear spin and electron spin.
The Zeeman frequencies of the electron and nuclear spin induced by the applied static magnetic field are denoted by $\omega_{e}$ and $\omega_{n}$, respectively.
We select two energy levels of the nuclear spin, $|\textnormal{-}1\rangle_{n}$ and $|0\rangle_{n}$, as the system, with the electron spin playing an ancillary role in the study, as illustrated in Fig.~\ref{fig1}.

To implement two-point measurement of work in the NV center system, we realized non-demolition projective measurement of the nuclear spin using the single-shot readout technique~\cite{Neumann2010}.
The photoluminescence rate of the NV center for $|0\rangle_e$ is higher than that for $|\textnormal{-}1\rangle_e$ with a contrast of about 30\%~\cite{Hrubesch2017}.
To perform projective measurement on the ${}^{14}$N nuclear spin in the computational basis, the electron spin was first polarized into $|0\rangle_e$, then flipped to $|\textnormal{-}1\rangle_e$ only if the nuclear spin state was $|\textnormal{-}1\rangle_n$ with a selective microwave pulse.
The selective microwave pulse was designed via optimal control to be robust against quasi-static noise.
Subsequently, a 532-nm laser pulse with a time duration of 200 ns was applied to excite the NV center and the fluorescence photons emitted were collected.
Our experiment was executed at a magnetic field of about 7600 G with its direction along the NV center symmetry axis.
This high external magnetic field can suppress the relaxation of nuclear spin, so the application of the 532-nm laser pulse almost does not alter the state of the nuclear spin.
By repeating the procedures above, we can accumulate the fluorescence signal to distinguish between different nuclear spin states.
Experimentally, the procedure was repeated 1200 times and the fidelity of the single-shot readout was $0.98(1)$.
Projective measurement along an arbitrary energy basis can be realized by applying appropriate rotations before the single-shot readout.

\begin{figure}
\centering
\includegraphics[width=\columnwidth]{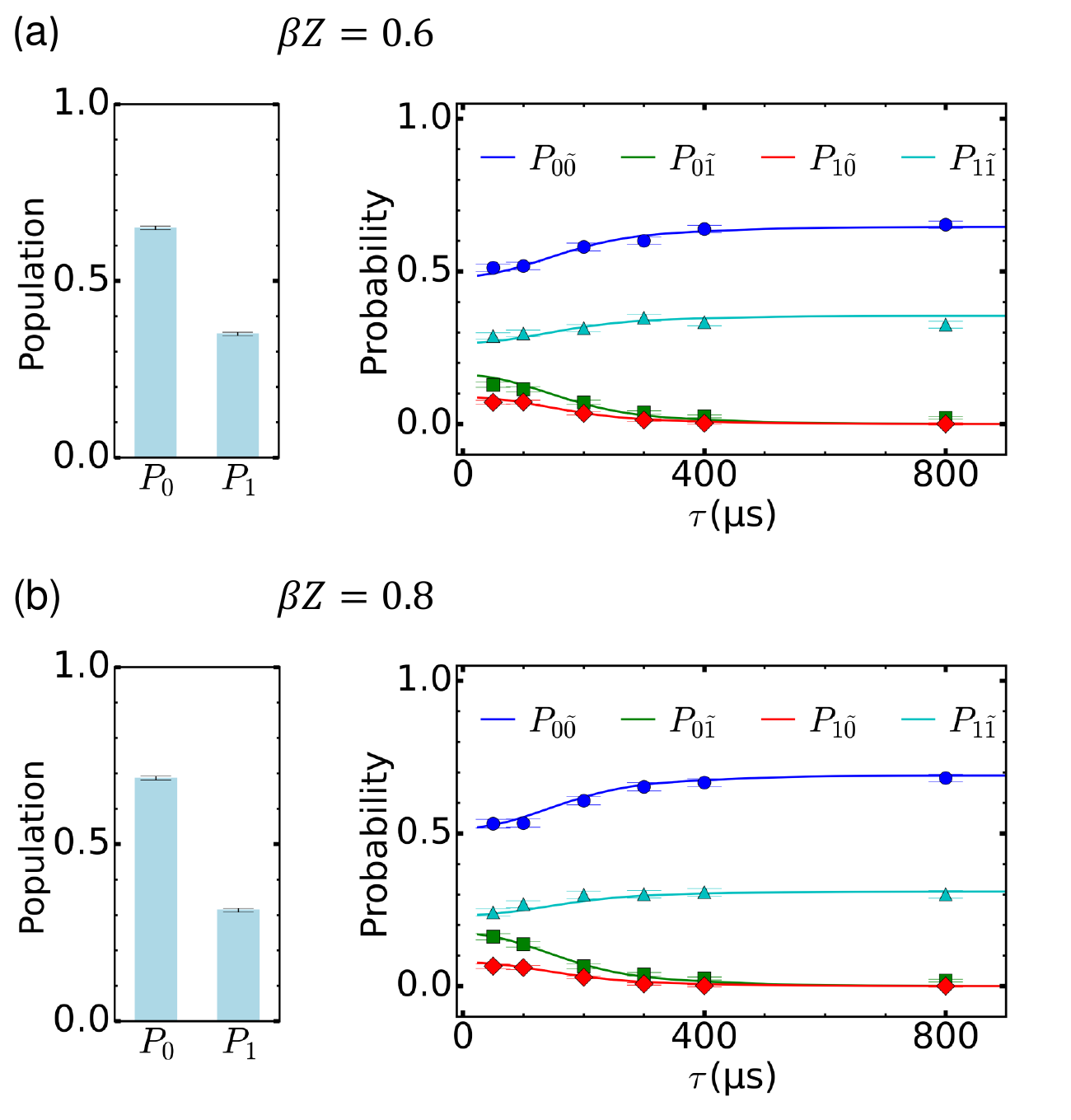}
\caption{\label{fig2} Thermal state and joint probability for effective temperature $\beta Z=0.6$ (a) and $\beta Z=0.8$ (b).
Left panels show the population of the initial thermal state. Right panels show the joint probability of different trajectories. Dots with error bars are experimental results. Pillars and lines show theoretical predictions.
}
\end{figure}

The pulse sequence to study the PMWF consists of four steps, as depicted in Fig.~\ref{fig1}. 
In our experiment, the switching of the system  Hamiltonian as a work protocol is chosen as follows:
\begin{equation}
H_0(t)=2\pi(ZS_z^{\prime}+X(t)S_x^{\prime})
\end{equation}
where $S_z^{\prime}=(|1\rangle\langle1|-|0\rangle\langle0|)/2$, $S_x^{\prime}=(|1\rangle\langle0|+|0\rangle\langle1|)/2$, $Z=5/\sqrt{3}~\rm{kHz}$, and $X(t)=5(1-\cos{(\pi t / \tau}))/2~\rm{kHz}$.
Here, energy levels $|0\rangle_{n}$ and $|\textnormal{-}1\rangle_{n}$ are relabeled with $|1\rangle$ and $|0\rangle$, respectively.
Firstly, we prepared the thermal state of the initial Hamiltonian $H(0)$ in three substeps.
Initially, the nuclear spin was initialized into state $|0\rangle$ through the single-shot readout and post-selection.
Next, a radiofrequency (RF) pulse $R_1$ was applied to prepare a coherent state $\sqrt{P_{\rm{thm}}^0}|0\rangle + \sqrt{P_{\rm{thm}}^1}|1\rangle$, where $P_{\rm{thm}}^0$ and $P_{\rm{thm}}^1$ are populations of the thermal state.
Then, the coherence of the nuclear spin was dissipated by applying two selective microwave (MW) pulses separated by a free evolution time being 10 $\upmu$s.
The selective MW pulse was applied to flip the electron spin from $|0\rangle_e$ to $|\textnormal{-}1\rangle_e$ conditioned that the nuclear spin state is $|0\rangle$.
The fidelity of the prepared thermal state in this way exceeds 0.9999 in our experiment.
After the thermal state was prepared, the first projective measurement was performed, projecting the nuclear spin onto an eigenstate of $H_0(0)$, such as $|E_m(0)\rangle$, with a probability of $p_m^0$. 
Subsequently, an RF pulse was applied to tune the Hamiltonian of the nuclear spin from $H_0(0)$ to $H_0(\tau)$ to realize the work protocol. Experimentally, the Rabi frequency and time duration of the RF pulse were adjusted to realize the work protocol with different adiabaticity. 
Finally, the second projective measurement was performed to project the nuclear spin onto an eigenstate of $H_0(\tau)$, such as $|\tilde{E}_{n}\rangle$, and gave the transition probability, $p^{\tau}_{m\rightarrow n}$.
Note that, the energy basis of the final Hamiltonian $H(\tau)$ differs from the computational basis, so an RF pulse $R_2$ was applied before the single-shot readout.

\begin{figure}
\centering
\includegraphics[width=\columnwidth]{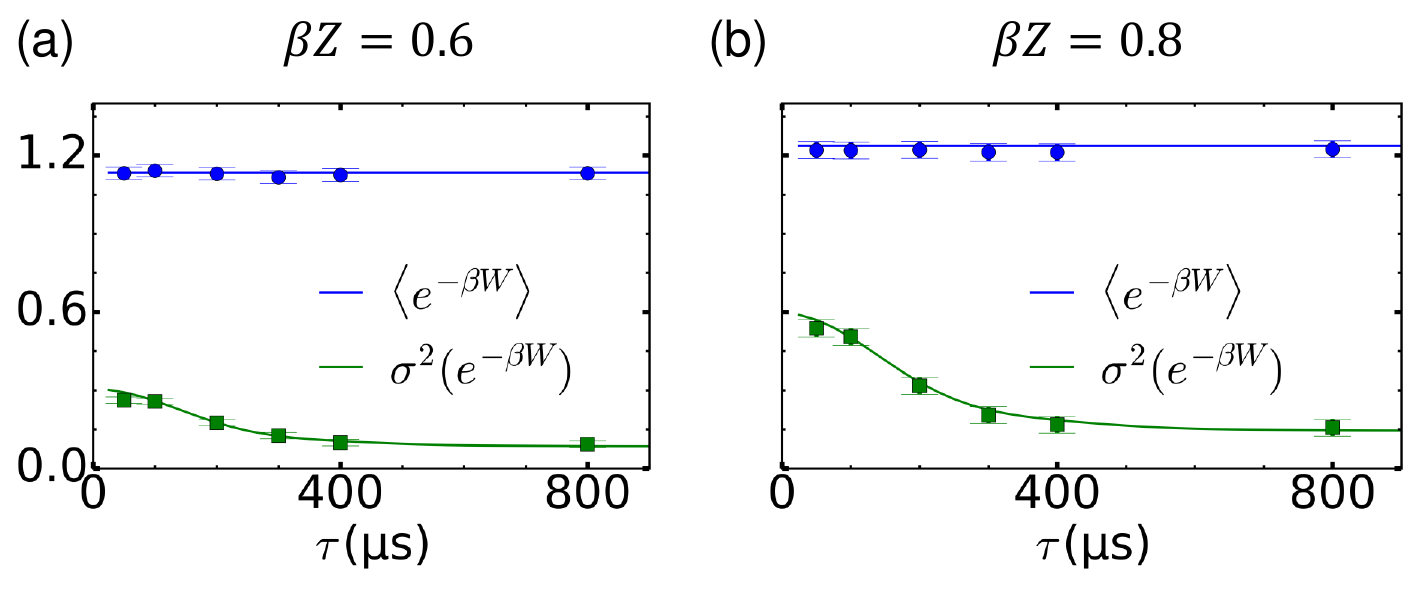}
\caption{\label{fig3}
Mean and variance of the exponential work versus the time durations of the work protocol. Blue circle dots (with error bars) and lines depict the experimental and theoretical mean values [denoted as $\langle e^{-\beta W}\rangle$], respectively. Green square dots (with error bars) and lines show the experimental and theoretical variances [denoted as $\sigma^{2}(e^{-\beta W})$], respectively. Inverse temperatures were $\beta Z=0.6$ (a) and $\beta Z=0.8$ (b).
}
\end{figure}

The results of the thermal state preparation and the joint probabilities $P_{m\tilde{n}}\equiv p^0_{m}p^{\tau}_{m\rightarrow n}$ are presented in Fig.~\ref{fig2}. 
Two different initial states were prepared according to two different inverse temperatures: $\beta Z=$ 0.6 and 0.8.
The measured populations are shown in left panels of Fig.~\ref{fig2}.(a) and (b), respectively.
These populations indicate that inverse temperatures of experimental prepared thermal states are $\beta Z = 0.62(2)$ and $0.78(2)$, which is consistent with our intended state preparation.
We varied the time duration $\tau$ of the work protocol from 50 $\upmu$s to 800 $\upmu$s to investigate how the fluctuation of $e^{-\beta W}$ changes as the protocol approaches the adiabatic limit.
To characterize the adiabaticity of work protocols, the adiabatic parameter was introduced as $\Gamma=\min_{t\in[0,\tau]}|\langle E_1(t)|\partial H(t)/\partial t|E_2(t)\rangle|/(E_1(t)-E_2(t))^2$~\cite{Sakurai2014}.
Here $|E_{1}(t)\rangle$, $|E_{2}(t)\rangle$ are two instantaneous eigenstates of $H(t)$ with $E_1(t)$, $E_2(t)$ being the corresponding energies.
When $\Gamma$ is much smaller than 1, the work protocol can be regarded as adiabatic. 
In our experiment, $\Gamma$ decreases from 1.014 to 0.063 as $\tau$ increases.
Joint probabilities with different time durations are displayed in right panels of Fig.~\ref{fig2}.(a) and (b). 
These joint probabilities presented here have been corrected to consider the influence of the infidelity of the single-shot readout (see Appendix A). $P_{mn}$ with $m\ne n$ characterizes the probability of non-adiabatic transitions that happened during the work protocol.    
As seen in Fig.~\ref{fig2}~(b) and (d),  $P_{m\tilde{n}}$ with $m\ne n$ decays to zero as $\tau$ increases, which fulfills the anticipation of the quantum adiabatic theorem.

\begin{figure}
\centering
\includegraphics[width=.99\columnwidth]{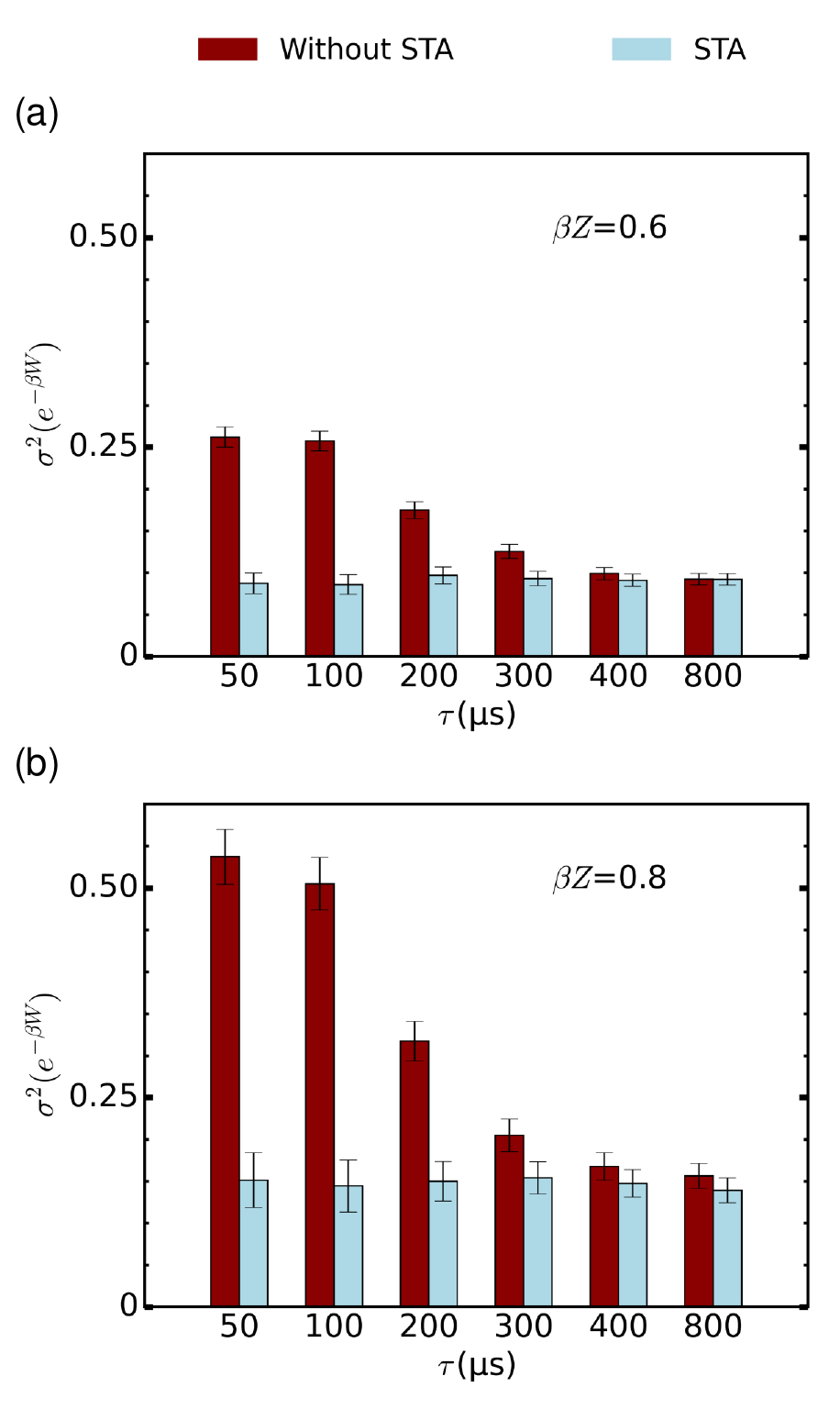}
\caption{\label{fig4} Variances of $e^{-\beta W}$ for normal and STA work protocols. Red (dark gray) and blue (light gray) bars show $\sigma^{2}(e^{-\beta W})$ of normal and STA work protocols with the same time duration, respectively. The minimal variance of $e^{-\beta W}$ can be achieved in fast work protocols assisted by STA. Inverse temperatures were $\beta Z=0.6$ (a) and $\beta Z=0.8$ (b).
}
\end{figure}

To investigate the PMWF, the average values and fluctuations of $e^{-\beta W}$ are experimentally obtained and displayed in Fig.~\ref{fig3}.
The mean values of $e^{-\beta W}$ for different time durations are plotted as blue circle dots in Fig.~\ref{fig3}~(a) and (b) for different values of $\beta$, with lines indicating the theoretical values of $e^{-\beta \Delta F}$.
All dots are consistent with lines when error bars were considered, verifying the validity of the Jarzynski equality.
The variance of $e^{-\beta W}$ calculated via Eq.~(\ref{work_fluctuation}) is presented by green square dots in Fig.~\ref{fig3}~(a) and (b).
The results clearly confirm that $\sigma^2(e^{-\beta W})$ decreases as $\tau$ increases. In particular,
it is observed that the variance of $e^{-\beta W}$ acquires its minimum value when the work protocol reaches adiabatic, as predicted by the PMWF.

Next we investigate the PMWF associated with fast work protocols. This is of great interest since an adiabatic process can be impractical in some situations, e.g., where inevitable decoherence becomes severe.   In an adiabatic process, the dynamics of the $n$th eigenstate follows the adiabatic path~\cite{Berry2009}, such that $|\psi_n(t)\rangle=e^{-i\int^t_0dt^{\prime}E_n(t^{\prime})-\int^t_0dt^{\prime}\langle E_n(t^{\prime})|\partial_{t^{\prime}}E_n(t^{\prime})\rangle}|E_n(t)\rangle$.
Here $|E_n(t)\rangle$ is the $n$th instantaneous eigenstate of $H_0(t)$.
To realize this evolution rapidly, a CD Hamiltonian is introduced such that the solution of the Schr{\"o}dinger equation $i\partial_t |\varphi(t)\rangle=H_{{\rm{CD}}}(t)|\varphi(t)\rangle$ is exactly $|\psi_n(t)\rangle$ given that $|\varphi(0)\rangle=|E_n(0)\rangle$.
In our model, the CD Hamiltonian takes the following form (see Appendix B),
\begin{equation}
H_{{\rm{CD}}}(t)= H_0(t)+Y(t)S_y^{\prime} ,
\end{equation}
where $Y(t) = Z\dot{X}(t)/(X^2(t)+Z^2)$ and $S_y^{\prime}=(-i|1\rangle\langle0|+i|0\rangle\langle1|)/2$.
The variance of $e^{-\beta W}$, denoted as $\sigma^2(e^{-\beta W})$, in such STA work protocols versus that in bare work protocols, both with inverse temperature $\beta Z=0.6$, is shown in Fig.~\ref{fig4}(a). 
The $\sigma^2(e^{-\beta W})$ in the STA work protocols are obviously less than those in bare and fast work protocols ($\tau=50,100,200,300~\upmu$s) without STA. 
When $\tau=800~\upmu$s, the bare work protocol can be regarded as adiabatic. The obtained variance of $e^{-\beta W}$ in the STA work protocols matches with that obtained in the bare and slow work protocol.
These results confirm that fast STA work protocols can exhibit the same minimal variance of $e^{-\beta W}$ as in an adiabatic work protocol.
This observation is further enhanced by considering a lower temperature in Fig.~\ref{fig4}(b) with $\beta Z=0.8$. It is also seen that the lower temperature case yields appreciably larger variance of $e^{-\beta W}$, indicating that for a fixed precision in estimating $\Delta F$, more experimental runs are required as temperature decreases to suppress the estimator bias.  

\section{IV. Conclusion}

Quantum effects on the statistics of $e^{-\beta W}$, a central quantity in the quantum Jarzynski equation, constitute an intriguing topic to understand the predictive power of the Jarzynski estimator in the quantum domain.  With the successful implementation of two-time measurement of quantum work, we have conducted a direct experimental investigation of the relationship between the variance of $e^{-\beta W}$ and adiabaticity of non-equilibrium work protocols.  We experimentally observed that adiabatic processes minimized the variance of $e^{-\beta W}$, and further studied the minimal variance of $e^{-\beta W}$ in fast work protocols assisted by shortcuts-to-adiabaticity control.  Our experimental results verify the so-called principle of minimal work fluctuations and shall stimulate future experimental studies on work fluctuations.    
For example, work protocols involving systems with infinite-dimensional Hilbert space tend to yield a diverging variance of $e^{-\beta W}$~\cite{Deng2017,Jaramillo2017}.
The convergence or divergence of the variance depends on many system parameters and can exhibit interesting phase diagrams~\cite{Jaramillo2017}.
The predictive power of the Jarzynski equality in such systems requires scrutiny.

\section{ACKNOWLEDGMENTS}

This work was supported by the National Key R\&D Program of China (Grants No. 2018YFA0306600 and No. 2016YFB0501603), the National Natural Science Foundation of China (Grants No. 12174373), the Chinese Academy of Sciences (Grants No. XDC07000000, No. GJJSTD2020000), Innovation Program for Quantum Science and Technology (Grant No. 2021ZD0302200), Anhui Initiative in Quantum Information Technologies (Grant No. AHY050000) and Hefei Comprehensive National Science
Center. X.R. thanks the Youth Innovation Promotion Association of Chinese Academy of Sciences for their support. W.L. is funded by Beijing University of Posts and Telecommunications Innovation Group.

W.C. and W.L. contributed equally to this work.

\section{APPENDIX A: Correction to joint probabilities}

In our experiments, we implemented the two-time measurement protocol to measure work statistics. 
In particular, we performed projective measurements at the start and the end of a switching process.
The required projective measurement was realized by single-shot readout and this procedure is not perfect due to errors mainly caused by the longitudinal relaxation process of the nuclear spin.\cite{Neumann2010}.
To mitigate the influence of such errors on our statistics, we considered a transition matrix (elaborated below) to correct the bare state-to-state transition probabilities accordingly.

Let us assume that the state before the readout is in a measurement basis $|j\rangle$.
Consider now a non-zero probability $p(i|j)$ that the output of the single-shot readout ends up with a different basis $|i\rangle$ (hence errors in our readout).
We can now define a transition matrix as $\boldsymbol{T}_{ij}=p(i|j)$.
Because our study involves a two-level system consisting of $|\textnormal{-}1\rangle_n$ and $|0\rangle_n$, the transition matrix is a two by two matrix:
\begin{equation}
\boldsymbol{T}=
\begin{bmatrix} 
p(\textnormal{-}1|\textnormal{-}1) & p(\textnormal{-}1|0) \\ 
p(0|\textnormal{-}1) & p(0|0)
\end{bmatrix},
\end{equation}
In our experiment, this transition matrix can be measured and was found to be
\begin{equation}
\boldsymbol{T}^{\rm{exp}}=
\begin{bmatrix} 
0.980 & 0.045 \\ 
0.020 & 0.955
\end{bmatrix},
\end{equation}
With this transition matrix that reflects our errors in state readout, we can correct our state-to-state transition probabilities.

First of all, the initial populations of our true prepared thermal state are $\boldsymbol{p}_0 = (p_0(-1),p_0(0))^T$, where $p_0(-1)$ and $p_0(0)$ represent the populations in states $|\textnormal{-}1\rangle_n$ and $|0\rangle_n$, respectively.
The measured populations are slightly different from $\boldsymbol{p}_0$ due to readout errors in the first projective measurement, resulting in $\boldsymbol{p}_0^{\rm{exp}}=\boldsymbol{T}^{\rm{exp}}\boldsymbol{p}_0$.
Thus, the corrected initial populations are given by $\boldsymbol{p}_0 = (\boldsymbol{T}^{\rm{exp}})^{-1}\boldsymbol{p}_0^{\rm{exp}}$.
Additionally, the readout error of the second projective measurement can also affect the measured state-to-state transition probabilities.
Suppose the true state-to-state transition probabilities due to the population transfer in our work protocol are given by,
\begin{equation}
\boldsymbol{P_c}=
\begin{bmatrix} 
p_c(\textnormal{-}1|\textnormal{-}1) & p_c(\textnormal{-}1|0) \\ 
p_c(0|\textnormal{-}1) & p_c(0|0)
\end{bmatrix},
\end{equation}
Then, the measured transtion probabilities become $\boldsymbol{P_c}^{\rm{exp}}=\boldsymbol{T}^{\rm{exp}}\boldsymbol{P}_c$.
This indicates that the corrected state-to-state transition probabilities are given by $\boldsymbol{P}_c = (\boldsymbol{T}^{\rm{exp}})^{-1}\boldsymbol{P}_c^{\rm{exp}}$.
With both the initial populations and the state-to-state transition probabilities corrected, the corrected joint probabilities for a sampled initial state making a transition to a final state are $(\boldsymbol{P})_{ij}=[(\boldsymbol{T}^{\rm{exp}})^{-1}\boldsymbol{P}_c]_{ij}[(\boldsymbol{T}^{\rm{exp}})^{-1}\boldsymbol{p}_0^{\rm{exp}}]_j$, which can be expressed as the following matrix:
\begin{equation}
\boldsymbol{P}=
\begin{bmatrix} 
p_c(\textnormal{-}1|\textnormal{-}1)p_0(-1) & p_c(\textnormal{-}1|0)p_0(0) \\ 
p_c(0|\textnormal{-}1)p_0(-1) & p_c(0|0)p_0(0)
\end{bmatrix}.
\end{equation}

\section{APPENDIX B: Counter-diabatic Hamiltonian to realize shortcuts-to-adiabatic}

\begin{figure}
\centering
\includegraphics[width=.8\columnwidth]{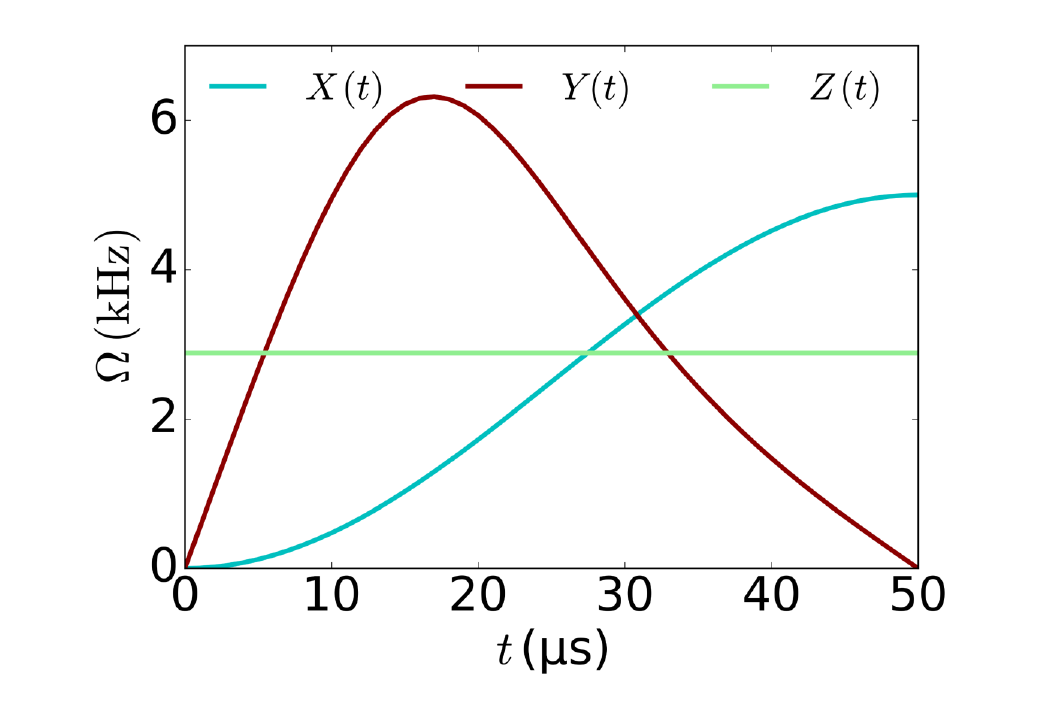}
\caption{\label{fig5} CD Hamiltonian to realize the STA process with switching time $\tau=50~\upmu$s.
}
\end{figure}

In our experiment, the chosen Hamiltonian in a bare work protocol takes the form as follows:
\begin{equation}
H_0(t)=2\pi(ZS_z^{\prime}+X(t)S_x^{\prime}).
\end{equation}
The instantaneous eigenstates of $H_0(t)$ is given by $H_0(t)|E_n(t)\rangle=E_n(t)|E_n(t)\rangle$.
In the adiabatic approximation, the dynamics of the $n$th eigenstate follows the adiabatic path, $|\psi_n(t)\rangle=e^{-i\int^t_0dt^{\prime}E_n(t^{\prime})-\int^t_0dt^{\prime}\langle E_n(t^{\prime})|\partial_{t^{\prime}}E_n(t^{\prime})\rangle}|E_n(t)\rangle$\cite{Berry2009}. Then the unitary operator is given by $U(t) = \sum_n e^{-i\int^t_0dt^{\prime}E_n(t^{\prime})-\int^t_0dt^{\prime}\langle E_n(t^{\prime})|\partial_{t^{\prime}}E_n(t^{\prime})\rangle}|E_n(t)\rangle\langle E_n(0)|$. To realize this unitary evolution, the counter-diabatic Hamiltonian is given by
\begin{equation}
\begin{aligned}
H_{\rm{CD}}(t)&=i\partial_tU(t)U^{\dagger}(t)\\
&=\sum_n E_n(t)|E_n(t)\rangle\langle E_n(t)|\\
&+i\sum_n |\partial_{t}E_n(t)\rangle\langle E_n(t)|\\
&-i\sum_n \langle E_n(t)|\partial_{t}E_n(t)\rangle|E_n(t)\rangle\langle E_n(t)|.
\end{aligned}
\end{equation}
The instantaneous eigenstates of $H_0(t)$ are $|E_0(t)\rangle=\cos(\theta(t)/2)|0\rangle+\sin(\theta(t)/2)|1\rangle$ and $|E_1(t)\rangle=\sin(\theta(t)/2)|0\rangle-\cos(\theta(t)/2)|1\rangle$, where $\theta(t)=\arctan(X(t)/Z)$. It is easy to calculate and find that $i\sum_n|\partial_{t}E_n(t)\rangle\langle E_n(t)|= \partial_t\theta(t)S_y^{\prime}=Z\dot{X}(t)/(X^2(t)+Z^2)S_y^{\prime}$ and $\langle E_n(t)|\partial_{t}E_n(t)\rangle = 0$ for $n=0,1$. 
Thus in our experiment, the counter-diabatic Hamiltonian takes the following form:
\begin{equation}
H_{\rm{CD}}(t)=2\pi(Z(t)S_z^{\prime}+X(t)S_x^{\prime}+Y(t)S_y^{\prime}),
\end{equation}
where $2\pi Y(t) = Z\dot{X}(t)/(X^2(t)+Z^2)$.
Taking the time duration $\tau=50~\upmu$s as an example, Fig.~\ref{fig5} depicts the time dependence of all specific components of $H_{\rm{CD}}(t)$.

\end{document}